\documentclass[aps,notitlepage,superscriptaddress,nofootinbib,pra,twocolumn,10pt]{revtex4-1}
  \usepackage{amsmath,amsfonts,amssymb,amsthm,graphicx,bbm,enumerate,times}
  \usepackage{mathtools}
  \usepackage[usenames,dvipsnames]{color}
  \usepackage[utf8]{inputenc}
  \usepackage{float}

  \usepackage{tikz}
  \usetikzlibrary{positioning,calc,arrows,shapes,fit,chains}
  \tikzset{join/.code=\tikzset{after node path={
  edge[every join]#1(\tikzchaincurrent)\fi}}}
  \tikzset{>=stealth', every join/.style={->}}
  \usepackage{pgfplots}
  \tikzset{
    spin/.style={inner sep=0,circle,draw=blue,
    fill=black,minimum height=0.5mm,minimum width=0.5mm}
  }

  \newtheorem*{lm*}{Lemma}

  \newtheorem*{prop*}{Proposition}

  \newtheorem*{co*}{Corollary}
  \newcommand{\Tr}{\text{Tr}}

  \newcommand{\G}{\mathcal{G}}
  \newcommand{\rel}{S}
  \DeclareMathOperator*{\argmax}{arg\,max}

  \newcommand{\fu}{Dahlem Center for Complex Quantum Systems, 
  Freie Universit\"{a}t Berlin, 14195 Berlin, Germany}

\begin{document}
  \title{Estimating strong correlations in optical lattices}

  \author{J. Gertis}
  \affiliation{\fu}
  \author{M. Friesdorf}
  \affiliation{\fu}
  \author{C. A. Riofr\'io}
  \affiliation{\fu}
  \author{J. Eisert}
  \affiliation{\fu}

  \begin{abstract}
    Ultra-cold atoms in optical lattices provide one of the most promising platforms for
    analog quantum simulations of complex quantum many-body systems.
	Large-size systems can now routinely be reached and are already used to probe a large variety of different physical
    situations, ranging from quantum phase transitions to artificial gauge theories. 
    At the same time, measurement techniques are still limited and full tomography
    for these systems seems out of reach. Motivated by this observation,
    we present a method to directly detect and quantify to what extent a quantum state deviates from a local
    Gaussian description, based on available noise correlation measurements from in-situ and time-of-flight measurements{. This is an} indicator {of the significance of} strong correlations in ground and thermal states, {as Gaussian states are precisely the ground and thermal states of non-interacting models}.
    We connect our findings, augmented by numerical tensor network simulations, to {notions of} equilibration, disordered systems and the suppression of transport in Anderson 
    insulators.
  \end{abstract}
  \maketitle

\subsection*{Introduction}
  Ultra-cold atoms in optical lattices provide one of the most prominent 
  architectures to probe the 
  physics of interacting many-body systems.
  The parameters of the Hamiltonians emerging in this fashion
  allow to explore a wide range of physical phenomena 
  in and out of equilibrium 
  \cite{BlochReview,Expansion,ChenSlow,Gauge}.
  They are one of the most promising platforms for realising 
  quantum simulations and show{ing} signatures of outperforming 
  classical computers for certain problems
  \cite{BlochSimulator,Trotzky12,MonteCarloValidator,Emergence}.
  Using state-of-the-art techniques, 
  large scale systems with several thousand atoms can be
  controlled \cite{BlochReview}; in fact, 
  even states with specific initial configurations and atoms aligned on
  largely arbitrary shapes can be realised \cite{1303.5652}.
  {Moreover, b}y modulating the optical lattice in time or by altering 
  its geometry, a wide range of complex physical
  settings can be explored, ranging from probing 
  quantum phase transitions \cite{MonteCarloValidator,Emergence} 
  to realising instances of artificial gauge theories \cite{Gauge}.

  While a large degree of control over these platforms has been achieved, 
  at present, it is still true that the 
  measurement capabilities are
  limited {in practice}. This appears particularly relevant in the context
  of quantum simulations, where the result of the simulation has to be read
	out from the physical experiment. 
  It seems clear that full quantum state tomography is infeasible,
  both for limits in {the availability of} measurement prescriptions as well as 
  due to the unfavourable scaling of the {tomographic }effort with the system size.
  Suitable combinations \cite{EfficientTomography} of tensor network tomography \cite{MPSTomo,cMPSTomographyShort} 
  and compressed sensing schemes \cite{Compressed} suggest a way forward towards
  achieving tomographic knowledge, but at present such ideas have not 
  been realised yet. 
  
  In the light of these obstacles, it seems imperative to 
  focus the attention to developing tools directly {that} detect relevant properties
  of the quantum state, rather than trying to capture the full density operator{---which gives} rise to information that is often not needed. Among those,
  entanglement features come to mind that contain valuable
  information about a quantum state \cite{quant-ph/0607167,Guehne,Audenaert06,1302.4897},
  or notions of non-classicality that can be directly detected \cite{NonClassical}. 
  Similarly, it is of interest to
  identify {and quantify} to what extent the state realised corresponds to a 
  ground or thermal state of an interacting model, and hence to what extent
  the state deviates from a Gaussian state. As one of the main
  promises of the field of ultra-cold atoms is to precisely study 
  interacting quantum many-body models deviating from non-interacting theories,
  this type of information is highly relevant, precisely probing to what extend an observation is compatible
  with a non-interacting model not exhibiting strong correlations.

  In this work, we introduce a new scheme that can be used to directly estimate 
  the local  Gaussianity of a state, based solely on 
  second and fourth moments of particle number measurements. 
  We apply this tool to the specific context of ultra-cold atoms 
  and show that noise correlations in in-situ measurements are already sufficient for its calculation. In this way, 
  we build upon and relate to the ideas of Refs.\ \cite{EhudCorrelations,NoiseInterferometry,Noise}, 
  but deliver an answer to the converse task: We do not show how interacting models are
  reflected in noise-correlations, but ask how data can be used to unambiguously witness
  such deviations from non-interacting models.

\subsection*{Gaussian states of massive particles}
  Interacting many-body quantum systems are exceedingly hard to capture and describe in
  terms of classical parameters. 
  Non-interacting models---models that have Gaussian ground and thermal states---are an exception to this rule, in that their
  description complexity is low.
  They are a paradigmatic class of states both for fundamental questions in quantum information
  as well as for finding ground states of condensed matter models,
  such as the interaction-free Bose-Hubbard model \cite{BlochReview}. {In this sense,} they are states that do not exhibit the intricate structure of interacting
  quantum many-body models. Again, since one of the main promises of the field is
  to address such interaction effects, it seems important to have tools at hand 
  to directly detect a deviation from Gaussianity.

  The bosonic Gaussian states discussed here are  
  characterised by the second moments collected in the correlation matrix $\gamma$ with entries
  \begin{align}
    \gamma_{i,j} = \Tr \, (b_i^\dagger b_j \, \rho) \; ,
  \end{align}
  where $b_j^\dagger, b_j$ denote the canonical bosonic creation and annihilation operators,
  $j=1,\dots, n$, {for an $n$ mode system.}
  Throughout this work, we investigate massive bosons leading to a situation in which 
  all $ \Tr \, (b_i b_j \, \rho) =0$ for all $i,j=1,\dots, n$.   Any such correlation matrix satisfying $\gamma\geq 0$ can be diagonalised
  with a unitary $V\in U(n)$ as $D= V\gamma V^\dagger$, reflecting a mode transformation 
  \begin{equation}
    b_j = \sum_{k=1}^n
    V_{j,k} \tilde b_k
  \end{equation}
  preserving the bosonic commutation relations{,  where $\tilde{b}_j$ are the transformed modes}. 
  One immediately finds 
     \begin{equation}
    \sigma = \argmax_{\substack{\rho \in \mathcal{D}\\ \Tr (\tilde{b}_j^\dagger \tilde{b}_k \rho) = \delta_{j,k} D_{k,k}}} S(\rho) ,
  \end{equation}
	where $S(\sigma) = - \Tr (\sigma \log \sigma)$ denotes the von-Neumann entropy and $\mathcal{D}$ denotes the set of density matrices. 
  In addition, by invoking the pinching inequality \cite{Bhatia-Matrix}, one finds that 
  it is already sufficient to fix the diagonal entries {of the correlation matrix} $D_{k,k}$ for $k=1,\dots, n$
  \begin{equation}
    \sigma = \argmax_{\substack{\rho \in \mathcal{D}\\ \Tr (\tilde{b}_k^\dagger \tilde{b}_k \rho) = D_{k,k}}} S(\rho).
  \end{equation}
  Following from this, given the diagonal elements of the correlation matrix in the
  momentum representation {of the modes}, the following Gaussian state is uncorrelated over the individual modes
  and given by (\ref{GGE})
  \begin{eqnarray} \label{GGE}
    \sigma = \prod_{k=1}^n \sigma_k,\,\, \sigma_k= \left( 1-e^{-\eta_k} \right) e^{- \eta_k \tilde{b}_k^\dag \tilde{b}_k},
  \end{eqnarray}
  where $\eta_k>0$ is corresponding to $D_{k,k}$ (see Appendix \ref{app_gaussian}).
  These states can be viewed as an instance of a generalised Gibbs ensemble \cite{RigolFirst,GeneralizedGibbs,CramerEisert}.
  Importantly for {the }context at hand, such Gaussian states also play a prominent role in the {setting} of optical lattices, 
  where they can, for example, be used to capture the superfluid ground state. 

\subsection*{{Deviation from} Gaussianity of states in optical lattices}
	In recent years, research on cold atoms in optical lattices has progressed
	significantly, by now allowing for unprecedented control of interacting
	quantum many-body system{s}, involving several thousands of atoms.  Relying on
	recent experimental advances, the position of individual atoms can be tracked
	using single site addressing \cite{Sherson-Nature-2010,Bakr-Science-2010}. 
	Using such techniques, local expectation
	values of the particle number as well as density-density correlations can be
	resolved. These local measurements have already provided important insight
	into the out-of-equilibrium dynamics of quantum many-body systems
	\cite{BlochReview} and allow{ed} to access their microscopic properties. {The main result of this work, which we detail subsequently, is} to identify tools to detect a deviation
	from Gaussianity---reflecting a non-interacting system---based on 
	{particle number measurements}, building upon Refs.\
	\cite{0805.1645,1008.4243}.  For this, we begin with a clarifying discussion
	to what extent states encountered in optical lattices can be Gaussian.

  States in optical lattices describe massive particles. For that reason, the particle number in
  each experimental run is fixed. This, however, implies that the full state is not Gaussian  {as we show below}.
  The best example for this is the perfect superfluid state, corresponding to the ground state of the
  1D free hopping Hamiltonian
  \begin{align}
    \label{eq:hop}
    H_{\mathrm{hop}} = \sum_{j=1}^{n-1} \left(b_j^\dagger b_{j+1} + b_{j+1}^\dagger b_j \right) .
  \end{align}
  Using a chemical potential $\mu>0$ to maintain the expected particle number {constant},
  thermal states of this Hamiltonian take the form
  \begin{align}
    \rho \propto e^{-\beta H - \mu \sum_{j=1}^n b_j^\dagger b_j} \; ,
  \end{align}
  where $\beta>0$ denotes the inverse temperature.
  This ensemble is Gaussian, as it is the exponential of a quadratic expression of creation and 
  annihilation operators.
  It, however, only fixes the particle number on average. 
  In order to fix the actual particle number, meaning to ensure that the
  state lives on a fixed particle number sector  
  also all higher moments have to be included
  in the ensemble.
  However, already including a fixed variance with a Lagrange parameter $\mu_2>0$
  \begin{align}
    \rho \propto e^{-\beta H - \mu \sum_{j=1}^n 
    b_j^\dagger b_j - \mu_2 (\sum_{j=1}^n b_j^\dagger b_j)^2} \; ,
  \end{align}
  results in a state that is no longer strictly Gaussian.

  Nevertheless, the superfluid {state} can be thought of being Gaussian in one important sense.
  Local particle number measurements are indistinguishable from a Gaussian state for sufficiently large 
  system sizes  is is not that obvious.  
  Thus, many measurements of the state can be captured using a simple Gaussian description.
  In the following, we present a general method that allows {us} to estimate the local {deviation from} Gaussianity of a
  state, based solely on measurements of second and fourth moments, which will afterwards be applied
  to states capturing optical lattices.

  \subsection*{Estimating the local {deviation from} Gaussianity}
  
    In order to {determine} the local  {deviation from} Gaussianity of states in optical lattices,
    we first describe how the distance of the global state to the manifold
    of Gaussian states can be captured and estimated relying only
    on second and fourth moments. We then use this insight to
    define a local  {deviation from} Gaussianity and evaluate it for paradigmatic bosonic models.

    To quantify the deviation of a state from Gaussianity, 
    we use the relative entropy as a natural quantity with a precise statistical interpretation 
    also known as Kullback-Leibler divergence.
    The relative entropy between two states $\rho$ and $\sigma$ is defined as
    \begin{align}
      \rel (\rho||\sigma) = \Tr(\rho \ln \rho) - \Tr(\rho \ln \sigma) \; .
    \end{align}
   This quantity provides the asymptotic statistical distinguishability of $\rho$ 
    from $\sigma$ in the situation of having {available} arbitrarily many copies of the state
    \cite{quant-ph/0102094,1106.1445v2}.

    Based on {the} relative entropy, we define the global  {deviation from} Gaussianity of a state as
    \begin{align}
      \label{eq:global_gauss}
      G (\rho) := \min_{\sigma \in \G} \rel (\rho||\sigma) \; ,
    \end{align}
    where $\G$ denotes the set of all Gaussian states. This can be seen as a measure of the 
    strong correlations present in the state, as quantifying the statistical deviation 
    from a state that could have been the ground or thermal state of a non-interacting model. 
    As it turns out, the minimum is always achieved for the Gaussian state $\sigma$ with the
    same correlation matrix {as $\rho$} \cite{1308.2939}.  
    We denote this special Gaussian state by $\sigma_\rho$. {The global}
     {deviation from} Gaussianity can be lower bounded by relying only on local measurement data,
    which will be performed in the following.

    We begin by using the additivity of the relative entropy which allows us to {describe the problem in} the symplectic eigenbasis
    of $\sigma_\rho$ and decompose the estimate into a problem involving individual eigen-modes \cite{quant-ph/0102094}
    \begin{align}
      G (\rho) = \rel (\rho||\sigma_{\rho}) 
      =& \rel \left( U(\gamma) \rho U(\gamma)^{\dag} || \otimes_{k=1}^n \sigma_k \right) \nonumber\\
      \geq& \sum_{k=1}^n \rel \left( \rho_k || \sigma_k \right) \; ,
    \end{align}
    where $U(\gamma)$ denotes the unitary transformation in Hilbert space (metaplectic representation) 
    reflecting the moment transformation into the 
    eigen-modes of $\gamma$, {$\rho_k := \Tr_{k^c} \left(U(\gamma) \rho U(\gamma)^{\dag} \right)$ and $\Tr_{k^c}$ denotes the reduction to the $k$-th mode.} 
    {Making use of the additivity of the relative entropy} for each individual mode, we can use the  {deviation from} Gaussianity of $\sigma_\rho$  
        and rewrite the estimate in terms of entropies \cite{0805.1645} 
    \begin{align}
      G (\rho) \geq& \sum_{k=1}^n \left( \rel (\sigma_k) 
      - \rel \left( \rho_k \right ) \right) \; .
    \end{align}
    These single mode entropy estimates are a major simplification compared to the original problem.

    Naturally calculating {or measuring} 
     the entropy of $\rho_k$ is still not a task that can be performed efficiently {or effectively}.
    Rather, we solely rely on {measured} fourth moments 
    $\Tr \, (\tilde{b}_k^\dagger \tilde{b}_k \tilde{b}_k^\dagger \tilde{b}_k \rho ) = M_{4,k}$
    and calculate the smallest possible distance to the manifold of Gaussian states compatible with this data.
    In this way, we are able to {compute} a minimal deviation from Gaussianity, thus showing that the
    state is, in this precise sense, strongly correlated.
    We therefore relax the problem as follows
    \begin{align}
      G (\rho) \geq& \sum_{k=1}^n \left( \rel (\sigma_k) - \rel \left( \kappa_k \right ) \right) ,\\ 
      \kappa_k :=& \argmax_{\substack{\kappa \in \mathcal{D}\\ 
      \Tr \, (\tilde{b}_k^\dagger \tilde{b}_k \kappa) = D_{k,k}\\
      \Tr \, (\tilde{b}_k^\dagger \tilde{b}_k \tilde{b}_k^\dagger \tilde{b}_k \kappa) = M_{4,k}} } S(\kappa) \; ,
    \end{align}
    where $\kappa_k$ is the maximum entropy state compatible with the second and fourth moments.      Using Schur's theorem \cite{Bhatia-Matrix},
    which states that the ordered eigenvalues of a matrix majorise the ordered diagonal entries
    \begin{equation}
      \mathrm{diag}(\rho) \prec \lambda(\rho) \; ,
    \end{equation}
    and the fact that the von-Neumann entropy is Schur convex, 
    the entropy of this state can be upper bounded by only considering the diagonal 
    \begin{equation}
      S(\rho_k) \leq \sum_i \rho_{k,i} \ln \rho_{k,i} \; ,
    \end{equation}
    {the second index labelling the main diagonal elements.}
        This allows for an efficient solution of the optimisation problem using Lagrange multipliers 
    (see Appendix \ref{app_singlemode}).
    Thus, we have seen that the global  {deviation from} Gaussianity defined in Eq.\ \eqref{eq:global_gauss} can
    be lower bounded by using second and fourth moments in the symplectic eigenbasis of the state.

    Based on this insight, we turn back to states describing massive particles in optical lattices.
    There, the full state cannot be reconstructed, which necessarily implies that {only} important features of the
    state {may be} addressed. 
    Here we focus on a particularly simple quantity that only relies on measuring the particle number
    on a single site, which is accessible experimentally using single-site addressing \cite{1303.5652}.
    Using such data, {we} define the \emph{local  {deviation from} Gaussianity} on site $j$ of a state as
    \begin{align}
      \label{eq:local_gauss}
      G_{\mathrm{local}}(\rho,j) :=& \argmax_{\substack{\kappa \in \mathcal{D}\\ 
      \Tr \, (n_j \kappa) = \bar{n}_j\\
      \Tr \, (n_j^2 \kappa) - \bar{n}_j^2 =  \overline{n_j^2} }} S(\kappa) \; ,
    \end{align}
    where $n_j$ is the particle number operator on site $j$ and $\bar{n}_j, \overline{n_j^2}$ denote 
    its experimentally measured expectation value and variance.

    This local  {deviation from} Gaussianity is not only {comparably} easily measurable, but more importantly also yields relevant
    information about the quantum system at hand.
    As discussed above, the superfluid state is locally Gaussian in the sense that the quantity
    in Eq.\ \eqref{eq:local_gauss} vanishes for large enough system sizes. 
    In this way, calculating the local  {deviation from} Gaussianity, which captures to what extend onsite particle measurements 
    are compatible with a Gaussian state, is a natural way to quantify the distance to a perfect superfluid state.
    Moreover, for the important case of non-interacting particles, it can also be used to identify the suppression
    of particle propagation due to disorder.
    Both these applications are elaborated upon below.

  \subsection*{Local  {deviation from} Gaussianity in the Bose-Hubbard model}
    In the following, we present numerical results for bosonic models commonly encountered in optical lattices.
    We begin with the attractive Bose-Hubbard model
    \begin{align}
      H_{\mathrm{BH}} = -\sum_{j=1}^{n-1} 
      \left(b_j^\dagger b_{j+1} + b_{j+1}^\dagger b_j \right) - \frac{U}{2} \sum_{j=1}^n 
      n_j (n_j-1) \; ,
    \end{align}
    where we have chosen the hopping strength equal to one and denote the interaction strength with $U$.
    The ground state for $U=0$ is the {idealised} superfluid state introduced above and thus locally Gaussian in our 
		sense, corresponding to a local  {deviation from} Gaussianity of zero.
    In contrast, when the interaction strength is increased, the system becomes strongly correlated
    and the size of the local  {deviation from} Gaussianity should increase.  
    
\begin{figure}
     \centerline {\includegraphics[width=.7\columnwidth]{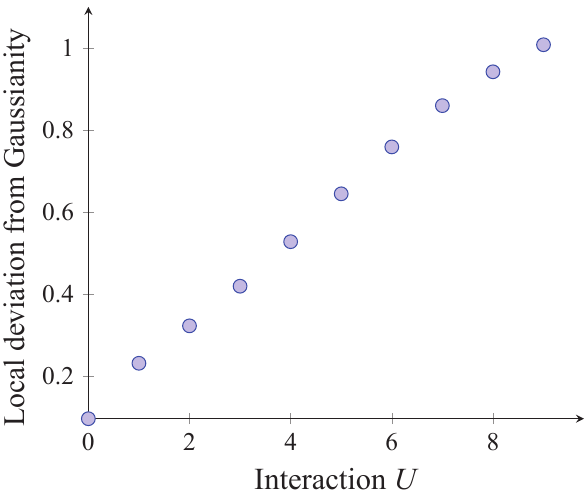}}
      \caption{Plotted is the local  {deviation from} Gaussianity based on the measurement of particle
        number and density-density correlator on a single lattice site of a 1D system 
        for the ground state of the attractive Bose-Hubbard model with filling fraction $\bar{n}=1$ 
        and interaction strength $U$.
        The results are obtained with exact diagonalisation using periodic boundary conditions 
        on $L=15$ lattice sites. 
      }
      \label{fig:ground}
    \end{figure}

    We have confirmed that this behaviour is indeed encountered numerically and find an almost linear
    {relation} between the local  {deviation from} Gaussianity and the interaction strength (see Fig. \ref{fig:ground}). In this way,
    one could even see the local  {deviation from} Gaussianity as an experimental probe to directly measure
    the interaction strength based solely on ground state particle number fluctuations.

    Another setting in which the local  {deviation from} Gaussianity is intriguing to investigate is the evolution
    of free systems, where it can be used as an indicator of disorder and the concomitant suppression
    of transport.
     
  \subsection*{Disordered systems}
    It is known  that non-interacting systems which exhibit transport in a suitable sense     evolve in time in a way that the states tend to locally Gaussian states
    following out of equilibrium dynamics \cite{1408.5148}, a feature that is true in surprising generality 
    \cite{CramerEisert,CramerCLT,Marek}.
     Thus, in this setting, the precise initial conditions are forgotten over time  and local expectation
    values can be captured using only the second moments of the initial state, then fully determining local
    expectation values.

    This applies in particular to the free hopping Hamiltonian introduced in Eq.\ \eqref{eq:hop}.
    A paradigmatic setting for this is given by an initial product state with one particle on every second site, 
    which can be experimentally prepared employing optical super-lattices \cite{Trotzky_etal12,BlochMBL}.
    This initial state is the ground state of an infinitely strongly interacting Bose-Hubbard model and
    it is thus far from being locally Gaussian, as described above.
    During time evolution that has transport, however, the particles distribute evenly over the lattice,
    thus moving towards the manifold of locally Gaussian states. 

\begin{figure}
      \centerline{\includegraphics[width=.7\columnwidth]{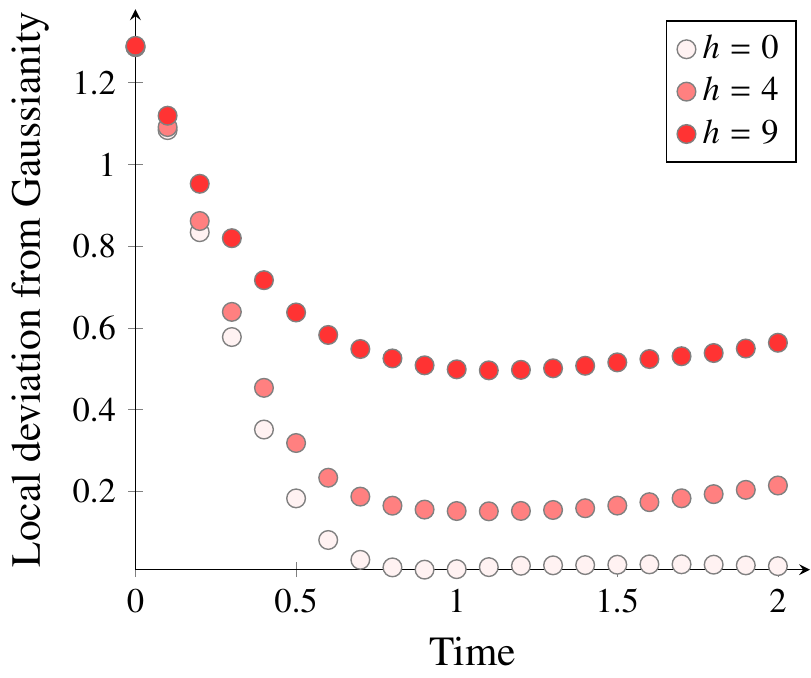}}
      \caption{Plotted is the time-dependence of the local  {deviation from} Gaussianity based on the measurement of particle
        number and density-density correlator on a single lattice site of a 1D system of $L=20$ sites. 
        The initial state is a charge-density wave corresponding to a Fock state with one atom 
        on every second lattice site and the evolution is governed by the Hamiltonian in Eq.\ \eqref{eq:H} with
        $w_j$ drawn randomly from the interval $[-h,h]$ for three different values of the disorder strength. 
        Shown is the average over $40$ disorder realisations.
        Initially, the local state is very non-Gaussian. 
        For the translationally invariant system showing transport, 
        information spreads through the lattice and the local measurements become compatible with a
        fully Gaussian description of the state, that is, the system dynamically gaussifies.
        In contrast, in the disordered case, transport is strongly suppressed and the non-Gaussianity
        remains locally visible.
      }
      \label{fig:time}
    \end{figure}

    For disordered systems, transport is strongly suppressed and the distribution of particles
    over the lattice thus does not take place.
    For concreteness, let us consider a simple 1D hopping model
    \begin{align}
      \label{eq:H}
      H = - \sum_{j=1}^{n-1} \left(b_{j}^\dagger b_{j+1} + b_{j+1}^\dagger b_{j} \right)
      + \sum_{j=1}^n w_j b_j^\dagger b_j \; ,
    \end{align}
    with local potentials $w_j$ drawn uniformly from some interval $[-h,h]$.
    For $h=0$ the model reduces to the free hopping and local Gaussification thus takes place.
    In contrast, when randomness is present, transport breaks down, resulting in a positive
    local  {deviation from} Gaussianity even for long times (see Fig. \ref{fig:time}).

\subsection* {{Deviation from} Gaussianity from time-of-flight}
  The same approach discussed here for local  {deviation from} Gaussianity can be applied to global properties of the lattice.
  In fact, it can be applied to any modes that are defined by a mode transformation
  \begin{equation}
   	\tilde{b}_q = \sum_{j=1}^n V_{q,j} b_j .
  \end{equation}
  In particular, it is applicable to time-of-flight measurements.
  If the quantum state is not translationally invariant due to the presence of a harmonic trap, an
  average  {deviation from} Gaussianity over the system size is then directly detected in this way.
  In time-of-flight measurements with finite accuracy of the camera pixels, effectively the diagonal elements 
  $\{\langle n(q) \rangle = \langle \tilde b_q^\dagger \tilde b_q\rangle\}$ of
  \begin{equation}
    \Gamma = V\gamma V^\dagger 
	\end{equation}
	are measured, more commonly expressed as
  \begin{equation}
    \langle n(q,t_{\text{ToF}}) \rangle = |\hat{w}_0(q)|^2 \sum_{j,k}
    e^{i q (r_j- r_k) - i 
    \frac{c (r_j^2 + r_k^2)}{t_{\text{ToF}}}} \langle b_j^\dagger b_k \rangle \;
  \end{equation}
  as a function of $q$,
  where $t_{\text{ToF}}$ is the time of flight and $c>0$ is a constant derived from the mass 
  and the lattice constant of the optical lattice. The fourth moments of the same 
  modes defined by $V$ are accessible as 
  $\{\langle \tilde{b}_q^\dagger \tilde{b}_q \tilde{b}_q^\dagger \tilde{b}_q\rangle\}$ and contained in the very same
  images from the laboratory, merely by computing higher moments, following a prescription of 
  Ref.\ \cite{EhudCorrelations}. 
  Let it be stressed again that in contrast to this reference,
  we aim for and provide a direct detection of correlations based solely on second and fourth moments.
  
  \subsection*{Outlook}
  
  In this work, we have introduced and elaborated on a method of directly detecting 
  local  {deviation from} Gaussianity of quantum many-body systems. In this way, we have identified a way of
  witnessing the deviation from the
  ground or thermal state of a non-interacting model, hence directly observing strong correlations present in the state.
  We did so by making use of ideas of convex optimisation. More conceptually speaking, the mindset
  that we advocated here is {to make use of currently available} tools to directly detect properties
  relevant for a research question at hand. {Given the simplicity of the bounds, }it constitutes 
  a relevant further step to equip the 
 bounds and estimates presented here with precise statistical confidence regions.  
   This complements a more conventional approach in which
  a situation  sharing some feature is modelled or classically simulated, and the predictions compared
  with those of data from measurements. It is the hope that the present work can contribute to the
  further development of such tools of certification for the study of many-body models, particularly
  relevant for the partial certification of the functioning and read out of analog quantum simulators.


\bibliographystyle{naturemag}

  \subsection*{Acknowledgements} We sincerely thank M.\ Ohliger and U.\ Schneider for stimulating discussions.
  We would like to thank the EU (AQuS, RAQUEL), the ERC (TAQ), the BMBF (Q.com), and the 
  DFG (EI 519/9-1, EI 519/7-1) for support. CAR also acknowledges the support of the Freie Universit{\"a}t 
  Berlin within the Excellence Initiative of the German Research Foundation.

\appendix
\section{Gaussian states and massive particles}
  \label{app_gaussian}
  In this appendix, we provide the details of a description of Gaussian states and
  their relation to states capturing massive particles.
  As described in the main text, the Gaussian state corresponding to measured second
  moments is given by 
  \begin{align}
    \sigma &= \argmax_{\substack{\rho \in \mathcal{D}\\ \Tr (\tilde{b}_k^\dagger \tilde{b}_k \rho) = D_{k,k}}} S(\rho) \; .
  \end{align}
  This maximum entropy state is achieved by the generalised Gibbs ensemble 
  \begin{align}
   \sigma&= \frac{1}{Z} \prod_{k=1}^n e^{-\eta_k \tilde{b}_k^\dag \tilde{b}_k} \; ,
  \end{align}
  where $Z$ is the usual partition sum and the $\eta_k$ are determined by demanding that 
  \begin{align}
    \Tr (\tilde{b}_k^\dagger \tilde{b}_k e^{-\eta_k \tilde{b}_k^\dag \tilde{b}_k} ) = D_{k,k} \; .
  \end{align}
  {This state is a Gaussian quantum state. The optimality can easily be seen by considering the relative entropy distance 
  \begin{equation}
  	S(\rho\|\sigma)  = -S(\rho) - {\rm tr}( \rho\log \sigma ), 
  \end{equation}
  where the term on the right hand side simplifies to $- {\rm tr}( \sigma\log \sigma )$ for a Gaussian state,
  	 and observing that this is a non-negative functional.}
  This expression can be calculated using the bosonic partition sum
  \begin{align}
    Z( \{\eta_k\}) &= \Tr \prod_{k=1}^n e^{-\eta_k \tilde{b}_k^\dag \tilde{b}_k} \\
                  &= \prod_{k=1}^n \sum_{j=0}^\infty e^{-\eta_k j}\\
                  &= \prod_{k=1}^n \frac{1}{1 - e^{-\eta_k}}.
  \end{align}
  With this, we straightforwardly haver
  
  \begin{align}
    D_{k,k} &= \frac{1}{Z} (-\partial_{\eta_k}) Z\\
            &= (1 - e^{-\eta_k}) \frac{e^{-\eta_k}}{(1 - e^{-\eta_k})^2}\\
            &= ({e^{\eta_k} - 1})^{-1}
              \end{align}
            and hence
     \begin{align}        
  \eta_k &= \ln(1+{D^{-1}_{k,k}}) \; .
  \end{align}

\section{Solving the single-mode problem}
  \label{app_singlemode}
  To conclude our estimate, we need to solve a convex minimisation
  problem subject to the measurement results
  \begin{align*}
    \text{minimize }& \quad \sum_n \rho_n \ln \rho_n \\
    \text{subject to}&~ \sum_n \rho_n = 1 ,\\
    &~\sum_n \rho_n n = M_2 ,\\
    &~\sum_n \rho_n n^2 = M_4 \; ,
  \end{align*}
  where, for convenience, we have suppressed the index $k$ labelling the mode in question.
  The Lagrangian to this problem is given by
  \begin{align*}
    \mathcal{L}(\rho,\mu_0, \mu_2, \mu_4) &= 
    \sum_k \rho_k \ln \rho_k + \mu_0(\sum_k \rho_k - 1) \\
    &+ \mu_2(\sum_k \rho_k k - M_2) + \mu_4(\sum_k \rho_k k^2 - M_4) 
  \end{align*}
  where $\mu_2,\mu_4$ are Lagrange multipliers.
  The extremal points of this Lagrangian are of the form $\rho_k = e^{-(1+\mu_0 + \mu_2 k +\mu_4 k^2)}$ for fixed $\mu_i$. 
  Thus the Lagrange dual function is given by
  \begin{align*}
    g(\mu_i)= &-\sum_k e^{-(1+\mu_0+\mu_2k+\mu_4k^{2})} -\mu_0 -\mu_2 M_2 -\mu_4 M_4 .
  \end{align*}
  {Invoking Lagrange duality, the bound of the main text can hence br proven.}

\end{document}